\documentclass[twocolumn,twoside,nofootinbib,showpacs,prd,aps,10pt]{revtex4}

\usepackage[hyperfootnotes=false,bookmarks=false]{hyperref}
\usepackage[dvips]{graphicx}
\usepackage[tbtags]{amsmath}
\usepackage{amssymb}

\renewcommand{\sec}[1]{Sec.~\ref{sec:#1}}

\newcommand{\eq}[1]{Eq.~\eqref{eq:#1}}

\newcommand{\nn}{\nonumber}

\newcommand{\Ecm}{E_\mathrm{cm}}
\newcommand{\zero}{{(0)}}

\newcommand{\df}{\mathrm{d}}

\newcommand{\bn}{\bar{n}}
\newcommand{\vn}{{\bf n}}
\newcommand{\op}{{\mathcal{O}}}

\newcommand{\al}{\alpha}
\newcommand{\ga}{\gamma}

\newcommand{\de}{\delta}
\newcommand{\eps}{\epsilon}
\newcommand{\si}{\sigma}

\newcommand{\hC}{\hat{C}}
\newcommand{\hO}{\hat{O}}

\begin{document}


\title{Comparison of leading and next-to-leading  logarithmic  electroweak corrections to Higgs production}

\author{Fabio Siringo}

\affiliation{Dipartimento di Fisica e Astronomia dell' Universit\`a di Catania and  INFN Sezione di Catania,\\
Via S.Sofia 64, I-95123 Catania, Italy.}

\date{\today}
\begin{abstract}
Using soft-collinear effective theory, the leading-log  radiative electroweak corrections are written in a closed and analytical form  for the hadronic cross section of Higgs production through vector boson fusion, 
$qq\to qqH$, one of the most promising channels for studying the Higgs boson at the LHC. 
The simple leading-log resummation is compared with a full next-to-leading-log calculation, and its accuracy
is found to be of order 1\% up to 10 TeV, i.e. better than the accuracy of PDFs. Corrections
are found to be larger than predicted by one-loop fixed order approximations at LHC energies.
The method provides a simple way of incorporating the electroweak corrections in software packages, 
improving the accuracy of simulations.
\end{abstract}



\maketitle 

\section{Introduction}

The recent discovery  of a new boson at  the Large Hadron Collider (LHC)\cite{LHC1,LHC2} 
will lead to a detailed study of its properties in the next years, and will require a careful comparison of the experimental data with the results of precise calculations for the Higgs sector of the Standard Model (SM).
Vector boson fusion (VBF), $qq\to qqH$, is the second largest channel for detecting the Higgs boson. It is a pure electroweak process and its study is very important for determining the couplings and for a deeper knowledge of the Higgs sector\cite{Duhrssen:2004cv,Zeppenfeld:2000td}. 
However, at the 10 Tev energy scale  of LHC,  even the electroweak radiative corrections become important~\cite{Ciccolini:2007jr,Ciccolini:2007ec} and should be included in computer simulations together with QCD corrections~\cite{Han:1992hr,Figy:2003nv,Berger:2004pca,Arnold:2008rz,Harlander:2008xn, Bolzoni:2010xr,Bolzoni:2010as,Bolzoni:2011cu}.
With a partonic center-of-mass energy $\sqrt{s}$
of several TeV - more than an order of magnitude larger than the masses $M_{W,Z}$ of the gauge bosons -
the radiative corrections contain large Sudakov logarithmic terms $\alpha^n L^m$ where 
$\alpha=\alpha_{1,2}$ are the weak coupling constants and the logarithms $L= \log  (s/M^2_{W,Z})$ 
emerge from the  two different energy scales.  
These terms dominate the perturbative expansion and may even require resummation when the fixed
order perturbation expansion breaks down.
However, for VBF the scattering amplitude is proportional to the vacuum expectation value (VEV),
and standard resummation methods do not apply because the effective operator is not a gauge singlet.  That makes VBF a special interesting process to deal with.

The Sudakov logarithms can be regarded as infrared logarithms since they diverge as $M_{W,Z}\to 0$, and by
using an effective theory they can be converted to ultraviolet logarithms and summed by standard
renormalization group techniques. Quite recently the soft-collinear effective theory  
(SCET) \cite{BFL,SCET1,SCET2,BPS}
 has been shown
to provide a simple way to obtain the sum of the series of leading-logs (LL) $\alpha^n L^{n+1}$,
next-to-leading-logs (NLL) $\alpha^n L^n$, next-to-next-to-leading-logs (NNLL) $\alpha^n L^{n-1}$, 
etc. ~\cite{Chiu:2007dg,Chiu:2007yn,Chiu:2008vv,Chiu:2009mg}. In the effective theory the single terms contributing to the  scattering amplitude are
multiplied by the general factor~\cite{Chiu:2008vv}
\begin{align} \label{amp}
{\cal  U}&(\mu_l,\mu_h) =\exp\left [ D_0(\alpha(\mu_l))+D_1(\alpha(\mu_l)) \log \frac{\mu^2_h}{\mu^2_l}\right] \nn \\ 
& \quad \times  \exp\left\{-\int_{\mu_l}^{\mu_h} \frac{ {\rm d}\mu}{\mu}
\left[ A(\alpha(\mu))\log\frac{\mu^2}{\mu^2_h}+B (\alpha(\mu))  \right]  \right\}   \nn \\ 
 &\quad \times  \exp F (\alpha(\mu_h))
\end{align}
where $\mu_l\approx M_Z$ is the low energy scale and $\mu_h\approx\sqrt{s}$ is the high energy scale.
The coefficient $F$ is the high scale matching coefficient, $D_0$ and $D_1$ are the low scale matching
coefficients, $B$ is the non-cusp anomalous dimension and $A$ is the coefficient of the cusp anomalous dimension.
The LL series is summed by the one-loop cusp anomalous dimension; the sum of the NLL series requires
the two-loop cusp anomalous dimension, the one-loop non-cusp anomalous dimension, and the one-loop low
scale matching coefficient $D_1$; the NNLL series is given by the the three-loop cusp,  two-loop $B$ and $D_1$, one-loop $D_0$ and $F$, etc.
In fact the exponentiated form of Eq.(\ref{amp})  only requires the inclusion of electroweak
corrections at low orders, while the unexponentiated form of fixed-order calculations would require
electroweak corrections of any order for achieving the same accuracy.
In some recent papers ~\cite{Fuhrer:2010vi, Siringo:2012} the method as been used for calculating the electroweak corrections to Higgs production via VBF. Numerical results at NLL order were obtained for the cross section, including the
effect of parton distribution functions (PDFs) ~\cite{Siringo:2012}.  Most of the corrections were obtained in analytical form by SCET and might be easily included in the other software packages that have been developed, without the need of tedious loop calculations. 

While PDFs imply a 3\% error, many of the retained
terms at NLL order are very small, less than 1\%, and could be neglected in order to speed up the numerical
integration of the cross section. 
In this paper the simpler LL order resummation is compared with the full NLL calculation for Higgs
production via VBF, and the accuracy of the two approximations is discussed. The LL calculation of the cross section is found to deviate from the NLL result by less than $1\%$ up to 10 TeV center-of-mass energy.
While the present study gives a useful evaluation of the accuracy in the SCET  resummation,
it provides a very simple and fast analytical way for including the electroweak radiative corrections in software packages.

In fact at NLL order the one-loop anomalous dimension turns out to
be a $10\times 10$ matrix in the operator basis~\cite{Siringo:2012}, and the running integral in Eq.(\ref{amp}) requires
a numerical computation. Moreover  two-loop beta functions are required for the running of
the couplings at the same order, yielding coupled equations that again must be solved by a numerical
routine. On the other hand,
at LL order the only one-loop term that is required is the cusp anomalous dimension, and by use of
the simple  uncoupled one-loop beta functions the running integral in Eq.(\ref{amp}) is analytical,
yielding a diagonal correction factor for the differential cross section.
Moreover, an exact cancellation of terms yields the same correction factor that would be found for
the quark scattering $qq\to qq$, plus Higgs rescattering and wave function renormalization terms.
NLL terms are still shown to be relevant for a full description of the dependence on scattering angles
of the differential cross section, but such small dependence is averaged in the integrated cross
section.

The paper is organized as follows: in \sec{kinematics}  the kinematics of the VBF process is described and some details on the integration of the cross section are reported; the general calculation framework is
described in \sec{operators} where the operator basis set is defined for the effective theory and matched
onto the low scale and high scale gauge theories; in \sec{running} the running of the Wilson coefficients
is discussed and the anomalous dimension is shown to take a simple diagonal form at LL order;
explicit analytical expressions are derived for the running integral in \sec{LLorder} and the numerical results
at LL order and  NLL order ar compared with other fixed order perturbative calculations in \sec{numerical}.

\section{Kinematics of VBF and cross section}
\label{sec:kinematics}

\begin{figure}[b] \label{fig:tree}
\centering
\includegraphics[width=0.48\textwidth]{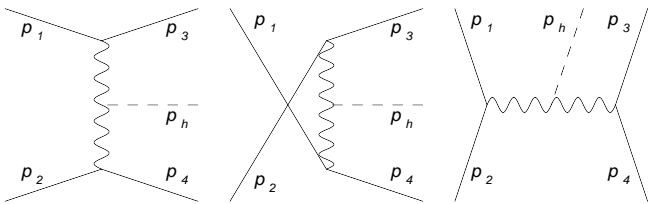}
\caption{Diagrams for electroweak Higgs production at tree-level:  $t-$, $u-$ and $s$-channel diagrams are reported.}
\end{figure}

At tree-level,  the Feynman diagrams for the process $qq \to qqH$ are shown in Fig.1. The two outgoing jets have a  large rapidity gap that characterizes the VBF channel, and the background is usually suppressed by cuts (for VBF cuts see e.g.~Refs.~\cite{Cahn:1986zv,Barger:1994zq,Figy:2003nv,DelDuca:2006hk}). 

We denote by $p_1$, $p_2$ the momenta of the incoming quarks, and by $p_3$, $p_4$
the momenta of the outgoing quarks (jets). 
The momentum of the Higgs boson is denoted by $p_h$,  and it is assumed to be on-shell, $p_h^2 = M_h^2$. 
Quark masses are neglected, $p_1^2 = p_2^2 = p_3^2 = p_4^2 = 0$, and all momenta are taken to be incoming, according to the conventions of Ref.~\cite{Fuhrer:2010vi}.
The kinematic can be expressed in terms of generalized Mandelstam variables:
\begin{align} \label{eq:mandelst}
s&=(p_1+p_2)^2 \,, \quad
t_3=(p_1+p_3)^2 \,, \quad
t_4=(p_2+p_4)^2 \,, \nn \\
u_3&=(p_2+p_3)^2 \,, \quad
u_4=(p_1+p_4)^2 \,.
\end{align}

In SCET,
a set of light-cone vectors is associated to each collinear direction. 
They are defined as
$n_i = \pm(1,\vn_i)$, $\bn_i = \pm (1,-\vn_i)$,
with the plus (minus) sign for incoming (outgoing) particles. 
For the quarks $n_i = \pm p_i/p_i^0$ while 
for the Higgs  $\vn_h = -{\bf p}_h/\sqrt{(p_h^0)^2 - M_h^2}$.

The colliding quarks ($i=1,2$)  carry 
a fraction $x_i$ of the total hadron momenta,
\begin{equation}
  p_i^\mu = x_i \Ecm \frac{n_i^\mu}{2}\,, \quad
  n_i^\mu = (1,0,0,\pm 1)\,, 
\end{equation}
where $\Ecm$ is the center-of-mass energy, and the fractions $x_i$ are integrated over in the cross section by use of PDFs that describe their distribution in the proton.
The momenta of the outgoing jets can be written as
\begin{align} \label{eq:p_34}
  p_3 &= -E_3 (1,\sin \theta_3,0,\cos \theta_3)\,, \nn \\
  p_4 &= -E_4 (1,\sin \theta_4 \cos \varphi, \sin \theta_4 \sin \varphi,\cos \theta_4)\,, 
\end{align}
where $E_i>0$ are the energies and $\theta_i$ the angles with the axis of the beam,
while $\varphi$ is the azimuthal angle between the outgoing quarks.   
The momentum of the Higgs is fixed by momentum conservation.

In the  effective theory, in order to evaluate the cross section, we square the matrix elements of the effective operators, 
sum over  helicities, flavors, channels and integrate over phase space:
\begin{align} \label{eq:si_ew}
\si_\mathrm{EW} &= \frac{1}{2\Ecm^2} \int \frac{\df x_1}{x_1}\, \frac{\df x_2}{x_2}\, \df \Phi_3\, \frac{1}{4} \sum_\mathrm{hel.} \sum_\mathrm{flav.} \sum_\mathrm{chan.} \nn \\ & \quad \times
\sum_{X,Y} f_i(x_1) f_j(x_2)\, \hC_X^* \hC_Y \langle \hO_X \rangle^* \langle \hO_Y \rangle
\,,\end{align}
where $\hO_X$ are the effective operators and $f_i(x)$ are the PDF for the flavor $i$  at momentum fraction $x$, and a factor of 1/2  must be included for identical particles in symmetric phase-space integrations. 
The Wilson coefficients $\hC$ are obtained in terms of tree-level high scale 
coefficients $C$ by matching at the high scale, running down to the low scale and  matching at the low scale
according to Eq.(\ref{amp}),
\begin{equation}\label{UC}
 \hC(\mu_l) = \mathcal{U}(\mu_l,\mu_h)\cdot C(\mu_h)
 \,
\end{equation}
which must be regarded as a matrix product in the operator basis set.
The three-body phase space $\df \Phi_3$  reads\cite{Siringo:2012} 
\begin{align} \label{eq:ph_sp}
  \int \df \Phi_3 & = \int\! \prod_{i=3,4} \frac{\df^3 p_i}{(2\pi)^3 2E_i} \frac{\df^4 p_h}{(2\pi)^3} \theta(-p_h^0) \de(p_h^2 - M_h^2) \nn \\ & \quad \times
 (2\pi)^4 \de^4(p_1+p_2+p_3+p_4+p_h) \nn \\ 
  & = \frac{1}{2^6 \pi^4} \int \prod_{i=3,4} \big[E_i \df E_i\, \df (\cos \theta_i)\big] \, \df \varphi\, \theta\bigg(\sum_{j=1}^4 p_j^0\bigg) \nn \\ & \quad \times
 \de\bigg[\bigg(\sum_{j=1}^4 p_j\bigg)^2 - M_h^2\bigg] \nn \\
 & = \frac{1}{2^6 \pi^4} \int\! \df (\cos \theta_3)\, \df (\cos \theta_4)\, \df \varphi\, \df E_3\,
  \frac{E_3 f_1}{f_2^2}
\,.
\end{align}
where
\begin{align}
  f_1 &= x_1 x_2 \Ecm^2 - M_h^2 - [x_1+x_2 + (x_2-x_1) \cos \theta_3] \Ecm E_3
  \nn \\
  f_2 &= [x_1+x_2 + (x_2 - x_1)\cos \theta_4] \Ecm
    -2[1-\cos \theta_3 \cos \theta_4 
  \nn \\ & \quad  
    - \cos \varphi \sin \theta_3 \sin \theta_4] E_3\,.
\end{align}
The remaining integrals in \eq{ph_sp} must be carried out numerically with
the boundary conditions 
\begin{align}
  &0 \leq \theta_3, \theta_4 \leq \pi
  \,, \qquad
  0 \leq \varphi \leq 2\pi
  \,, \nn \\
  &\frac{M_h^2}{\Ecm^2} \leq x_1 \leq 1
  \,, \qquad
  \frac{M_h^2}{x_1 \Ecm^2} \leq x_2 \leq 1  
  \,, \nn \\  
  &0 \leq E_3 \leq \frac{x_1 x_2 \Ecm^2 - M_h^2}{[x_1+x_2+(x_2-x_1) \cos \theta_3]\Ecm}
  \,,
\end{align}
While we already included the bounds of the PDFs, the boundaries can be restricted by further cuts that we might  impose on the phase space in the integration.

\section{Effective theory and Operator basis set}
\label{sec:operators}

Electroweak corrections can be obtained by SCET in the framework 
of Ref.~\cite{Chiu:2007dg,Chiu:2009mg}. The extension to the VBF process 
was derived in Ref.~\cite{Fuhrer:2010vi, Siringo:2012}, and the explicit expressions are
reported in Ref.~\cite{Siringo:2012}.
Here we only consider the SM gauge group. Extensions like the minimal left-right symmetric gauge
group\cite{Siringo:2004} will be the subject of an other paper.

The first step consists of matching onto SCET at a high scale $\mu_h \sim \sqrt{s}$. 
Here the effects of symmetry breaking are suppressed,
and the matching can be done in the unbroken  gauge theory.
Next we run the effective operators down to a low scale $\mu_l \sim M_Z$ by renormalization group equations. At the low-scale, the $W$ and $Z$ boson are integrated out: we match onto a $SU(3) \times U(1)$ effective theory, only containing gluons and photons. In this low-scale matching, the effects of $SU(2) \times U(1)$ symmetry breaking must be considered. They only
enter in this low-energy matching.
 
At the high scale, the operators for VBF are given by~\cite{Fuhrer:2010vi,Siringo:2012}
\begin{align} \label{eq:op_basis}
  \op_{1A,B} &= O_1 O_A\,, O_1 O_B\,, \nn \\
  \op_{2A,B,C} &= O_2^a O_A^a\,, O_2^a O_B^a\,, O_2^a O_C^a\,, \nn \\
  \op_{3A,B} &= O_3 O_A\,, O_3 O_B\,, \nn \\
  \op_{4A,B,C} &= O_4^a O_A^a\,, O_4^a O_B^a\,, O_4^a O_C^a\,.
\end{align}
In the Higgs sector we introduce the operators $O_1, \dots, O_4$
\begin{align} \label{eq:op_higgs}
O_1 &= \Phi^\dagger_h \phi_0 + \phi_0^\dagger \Phi_h\,, \nn \\
O_2^a &= \Phi^\dagger_h T^a \phi_0 - \phi_0^\dagger T^a \Phi_h\,, \nn \\
O_3 &= \Phi^\dagger_h \phi_0 - \phi_0^\dagger \Phi_h\,, \nn \\
O_4^a &= \Phi^\dagger_h T^a \phi_0 + \phi_0^\dagger T^a \Phi_h \,,
\end{align}
where $\Phi_h = W_{n_h}^\dagger \phi_{n_h}$  is the collinear scalar doublet $\phi_{n_h}$ that gives rise to a Higgs boson, with its collinear Wilson line $W_{n_h}$. 
The field $\phi_0$ is a soft scalar that gives rise to a VEV when the symmetry is broken. 
For the quarks we introduce the operators
\begin{align} \label{eq:op_ferm}
 O_A &= \bar \Psi_3 \ga^\mu T^a \Psi_1 \bar \Psi_4 \ga_\mu T^a \Psi_2 \,, \nn \\
 O_B &= C_F \bar \Psi_3 \ga^\mu \Psi_1 \bar \Psi_4 \ga_\mu \Psi_2 \,, \nn \\
 O^a_A &= \bar \Psi_3 \ga^\mu T^a \Psi_1 \bar \Psi_4 \ga_\mu \Psi_2 \,, \nn \\
 O^a_B &= \bar \Psi_3 \ga^\mu \Psi_1 \bar \Psi_4 \ga_\mu T^a \Psi_2 \,, \nn \\
 O^a_C &= i \eps^{abc}\, \bar \Psi_3 \ga^\mu T^b \Psi_1 \bar \Psi_4 \ga_\mu T^c\Psi_2
\,,
\end{align}
with the subscript $i=1,\dots,4$  that labels the momentum $p_i$ of the particle.  
Collinear Wilson lines $\Psi_i = W_{n_i}^\dagger \psi_{n_i}$ are included in all these fermion fields, 
according to collinear gauge invariance. In order to keep the notation as general as possible,
the projectors $P_{R,L} = (1 \pm \ga_5)/2$  have been suppressed, and both left- and right-handed quarks have been allowed. Of course the field $\Psi_i$ is supposed to be a fermion doublet (singlet) if it is left-handed (right-handed). 
 
Each helicity  is considered separately and the contributions are combined at the end.
It is quite obvious that the operators $\op_A$ and $\op_C^a$ can only make sense  if the quarks are left-handed, while the operators $\op_A^a$ and $\op_B^a$ are allowed if at least one of the incoming particles is left-handed. 

As discussed in Ref.\cite{Siringo:2012} this operator basis is not complete, because it only
suffices for quarks and for the $t$-channel. However for incoming anti-quarks and for the
$u$-channel the corresponding operators are obtained by interchanging the particle labels.
Thus we will only discuss  the case of incoming quarks in the $t$-channel. 

At LL and NLL order the tree-level high-scale matching suffices, and can be done in the unbroken phase of the electroweak gauge theory.  The coupling of the scalar doublet to the gauge fields
is described by the Lagrangian terms
\begin{align}
{\cal L}=&\,\frac{1}{4} \big(g_2^2 A^a_\mu A^{a\mu} + g_1^2 B_\mu B^\mu \big)\Phi^\dagger\Phi \nn \\
& + g_1g_2 B^\mu A^a_\mu\, \Phi^\dagger T^a \Phi \,.
\end{align}
In Fig. 1 we only need to consider the $t$-channel, because the $s$-channel contribution is suppressed in VBF. Matching onto the operators in \eq{op_basis}, yields
\begin{align} \label{eq:c_def}
  C_{1A}(\mu) &= \frac{i g_2^4(\mu)}{2t_3 t_4}\, \eta_1 \eta_2\,, \ \
  C_{1B}(\mu) = \frac{2i g_1^4(\mu)}{3t_3 t_4}\, Y_1 Y_2\,, \nn \\
  C_{4A}(\mu) &= \frac{i g_1^2(\mu) g_2^2(\mu)}{t_3 t_4}\, \eta_1 Y_2\,, \nn \\
  C_{4B}(\mu) &= \frac{i g_1^2(\mu) g_2^2(\mu)}{t_3 t_4}\, Y_1 \eta_2\,. 
\end{align}
while the other coefficients vanish at tree level. In order to keep the notation as general as
possible,  a new variable $\eta_i$ has been defined: $\eta_i=1$ if the particle with label $i$ is left-handed and $\eta_i=0$ if the particle is right-handed. 
The hypercharge is $Y=1/6$ for left-doublets, while for right-handed particles $Y=2/3$ for
up-type quarks and $Y=-1/3$  for down-type quarks. 

The running of the Wilson coefficients follows by the RG equation 
\begin{equation} \label{eq:RGE}
 \mu \frac{\df}{\df\mu}C(\mu) = \ga(\mu) C(\mu)
\,,\end{equation}
in terms of the  anomalous dimension $\ga$ of  the operators.
That is a matrix equation, 
corresponding to the 10 operators in \eq{op_basis}. Using the notation of Eq.(\ref{amp}) the anomalous dimension can be written as the sum of cusp and non-cusp terms 
\begin{equation} \label{anomal}
\ga(\mu)=  A(\alpha(\mu))\log\frac{\mu^2}{\mu^2_h}+B (\alpha(\mu))  
\,,\end{equation}
We only need the one-loop cusp anomalous dimension at LL order, while two-loop cusp  and one-loop
non-cusp terms were required at NLL order and were reported in detail in Ref.~\cite{Siringo:2012}.

Finally, let us consider the low energy matching.  The tree-level result suffices at LL order. The one-loop calculation was required at NLL order and was also reported in Ref.~\cite{Siringo:2012}.
At low energies we must take into account the effects of electroweak symmetry breaking.
We match onto a basis of operators in the broken phase of the gauge group. 

 For the Higgs part of the operator, given in \eq{op_higgs},  the soft scalar field simply attains a VEV and the collinear scalar field produces a Higgs.  Thus at tree-level the low energy matching
yields
\begin{equation} \label{eq:Higgs_tree}
  \{O_1,O_2^a,O_3,O_4^a\} \to \{1,0,0,-\de^{a3}/2\} v h\,
\end{equation}
where $v$ is the VEV and $h$ is the Higgs field. 

The quark part in \eq{op_ferm} gets matched onto the set
\begin{align} \label{eq:br_basis}
  \hO_A = \bar u_3 \ga^\mu u_1 \bar u_4 \ga_\mu u_2 \,, \nn \\
  \hO_B = \bar u_3 \ga^\mu u_1 \bar d_4 \ga_\mu d_2 \,, \nn \\
  \hO_C = \bar d_3 \ga^\mu d_1 \bar u_4 \ga_\mu u_2 \,, \nn \\
  \hO_D = \bar d_3 \ga^\mu d_1 \bar d_4 \ga_\mu d_2 \,, \nn \\
  \hO_E = \bar d_3 \ga^\mu u_1 \bar u_4 \ga_\mu d_2 \,, \nn \\
  \hO_F = \bar u_3 \ga^\mu d_1 \bar d_4 \ga_\mu u_2 \,.
\end{align}
Here $u$ and $d$ denote up and down-type quarks. 
For operators $\hO_A,\dots, \hO_D$, the  pairs of fields $\bar \psi \ga^\mu \psi$ can be
 left-handed or right-handed, whereas in $\hO_E$ and $\hO_F$ all fields must be left handed.
We match \{$\op_{1A,B}$, $\op_{4A,B,C}$\} onto the  operators $\{\hO_A,\dots, \hO_F\}$ in \eq{br_basis}, while we
ignore $O_2^a$ and $O_3$ which vanish at tree level. The matching  is described by the
$6 \times 5$ matrix $R^\zero$ which is defined according to
\begin{align}\label{eq:D0}
  O_i \to v h\sum_J \hO_j R^\zero_{ji}\,, \quad
  \hC_j = v\sum_i R^\zero_{ji}C_i 
\,,\end{align}
so that
\begin{equation}
  \sum_i C_i O_i \to \sum_i \hC_i h \hO_i 
\,.\end{equation}
The matrix $R^\zero$ follows
\begin{equation}
R^\zero = \frac{1}{4}
\begin{pmatrix}
1 & 3 & -1 & -1 & 0\\
-1 & 3 & -1 & 1 & 0\\
-1 & 3 & 1 & -1 & 0\\
1 & 3 & 1 & 1 & 0\\
2 & 0 & 0 & 0 & -1 \\
2 & 0 & 0 & 0 & 1
\end{pmatrix}\,.
\end{equation}

\section{Anomalous dimension and running}
\label{sec:running}

At LL order, one-loop corrections enter through the cusp anomalous dimension. For VBF the one-loop
terms contributing to the anomalous dimension were derived in Ref.~\cite{Fuhrer:2010vi}  and reported in
detail in Ref.~\cite{Siringo:2012}. They can be written as the sum of three terms
\begin{equation}\label{eq:ga_total}
\ga = \ga_C +\ga_S+ \hat \ga. 
\end{equation}

The first term $\ga_C$ is the collinear anomalous dimension, it is diagonal and contains the large logarithms
that contribute to the cusp anomalous dimension, plus wave function renormalization terms. 
Resummation at LL order requires the inclusion of these one-loop terms, and they are easily obtained by the sum of single-particle collinear terms~\cite{Chiu:2009mg,Chiu:2009ft} for all  the external particles 
\begin{equation}\label{eq:ga_C}
\ga_C = \sum_i \ga^i_C +\ga^h_C
\end{equation}
As reported in Ref.~\cite{Siringo:2012}, with the shorthand
\begin{equation}
L_i = \log \frac{\bn_i \cdot p_i}{\mu} \, ,
\end{equation}
the quark one-loop  collinear terms $\ga^i_C$  read
\begin{equation}
  \ga_C^i = \Big[\frac{3\al_2}{4\pi} \eta_i + \frac{\al_1}{\pi} Y_i^2\Big] L_i 
- \frac{9 \al_2}{16\pi} \eta_i - \frac{3 \al_1}{4\pi} Y_i^2\,, 
\end{equation}
while the Higgs one-loop  collinear term $\ga^h_C$ is
\begin{equation}
\ga_C^h = \Big[\frac{3\al_2}{4\pi}  + \frac{\al_1}{4\pi}  \Big] L_h 
- \frac{3 \al_2}{4\pi} - \frac{\al_1}{4\pi} + \frac{3y_t^2}{16\pi^2}\,,
\end{equation}
where we included the contribution from the top Yukawa $y_t$ to the Higgs wave function renormalization.

The second term $\ga_S$ in Eq.(\ref{eq:ga_total}) is the soft anomalous dimension that is obtained by
summing over the soft functions~\cite{Chiu:2009mg,Chiu:2009ft}, and is reported in Ref.~\cite{Siringo:2012} in some detail. This soft term does not contain any large logarithm and does
not contribute to the cusp anomalous dimension. It only depends on the scattering angles of the external particles, and its effect largely cancels in the integration of the cross section. At LL order this term
can be neglected.

The third term  $\hat \ga$ in Eq.(\ref{eq:ga_total}) is a specific new contribution occurring in the VBF process~\cite{Fuhrer:2010vi}, and can be written as the sum of $SU(2)$ and  $U(1)$ parts plus a
term $\hat \ga^\lambda$ arising from the rescattering  of the Higgs boson. The first two parts contain
cusp diagonal terms proportional to $\log \mu$, and smaller non-cusp off-diagonal terms that do not depend on the scale $\mu$.
At LL order we only need to retain  the diagonal cusp terms and the rescattering term that is non-cusp
but  is large because of the large self-coupling $\lambda$ of the Higgs boson. We also include the diagonal wave function renormalization terms.  According to Ref.~\cite{Siringo:2012}, the cusp parts of the $SU(2)$ contribution read
\begin{equation} \label{eq:su2_1}
  \hat \ga^{SU(2)} _{1A,1B}=- \frac{3\al_2}{8\pi} [ 2L_h + 1]
\end{equation}
for the coefficients $C_{1A}$, $C_{1B}$,
\begin{equation} \label{eq:su2_4A}
  \hat \ga^{SU(2)} _{4A}=\frac{\al_2}{8\pi} [ 2L_h-4\eta_1(L_1+L_3) + 1]
\end{equation}
for the coefficient $C_{4A}$ and 
\begin{equation} \label{eq:su2_4B}
  \hat \ga^{SU(2)} _{4B}=\frac{\al_2}{8\pi} [ 2L_h-4\eta_2(L_2+L_4) + 1]
\end{equation}
for the coefficient $C_{4B}$.
The cusp part of the $U(1)$ contribution is
\begin{equation} \label{eq:u1}
  \hat \ga^{U(1)}=- \frac{\al_1}{8\pi} [ 2L_h + 1],
\end{equation}
and the Higgs rescattering contribution is given by
\begin{equation} \label{eq:lambda}
\hat \ga^\lambda=\frac{c^\op \lambda}{4\pi^2}
\end{equation}
where $c^\op = \{3,3,1,1\}$ for the coefficient  $\{C_{1A}, C_{1B}, C_{4A}, C_{4B} \}$. 

Summing all the terms, we find that the dependence on $L_h$ cancels (up to non-cusp terms like
the difference $ L_h-L_i$ that can be neglected at LL order), yielding the following simple
result for the total anomalous dimension
\begin{equation} \label{eq:ga_tot}
\ga = \frac{3\al_2}{4\pi} \sum_i \eta_i  L_i + \frac{\al_1}{\pi} \sum_i Y_i^2 L_i +\ga_{nc}
\end{equation}
where the sum is over the four external quarks. Here
 $\ga_{nc}$ is the sum of the retained diagonal non-cusp terms, including constant wave function renormalization and Higgs rescattering terms
\begin{equation} \label{eq:ga_nc}
\ga_{nc} = - \frac{9\al_2}{16\pi}\left[ \sum_i \eta_i  + d^\op \right]
-\frac{3\al_1}{4\pi}\left[ \sum_i Y_i^2 + \frac{1}{2}\right] +\frac{3y_t^2}{16\pi^2}+\hat \ga^\lambda
\end{equation}
where $d^\op = \{2,2,10/9,10/9\}$ for the coefficient  $\{C_{1A}, C_{1B}, C_{4A}, C_{4B} \}$.
Some of these terms are small, but they are constant, do not depend on scattering angles, and their
weight might sum up in the integration of the cross section. All other non-cusp terms  have been neglected.

The simple result of Eq.(\ref{eq:ga_tot}) says that the one-loop cusp anomalous dimension is the same that
we would obtain by the sum of the collinear terms for the external quarks, neglecting the Higgs particle.
However the Higgs momentum would affect the kinematic of the quarks anyway.  Moreover the Higgs  rescattering and wave function renormalization terms are not small and have been included in the non-cusp
part $\gamma_{nc}$ of the anomalous dimension.

\section{radiative corrections at LL order}
\label{sec:LLorder}

At LL order the running of the coupling constant can be evaluated by uncoupled one-loop beta functions.
That, together with the simple diagonal form of the one-loop cusp anomalous dimension in Eq.(\ref{eq:ga_tot}),
allows for a fully analytical evaluation of the electroweak radiative corrections.
 
By insertion of Eq.(\ref{eq:D0}) in Eq.(\ref{amp}), the low scale Wilson coefficients follow from Eq.(\ref{UC}) that
now reads
\begin{equation}\label{eq:WC}
  \hC_Y (\mu_l)= v\sum_Y R^\zero_{YX}
\left\{ \prod_{(\al)} {\cal A}_X^{(\al)} (\mu_l, \mu_h) \right\} C_X (\mu_h)
\end{equation}
where the product runs over the four couplings $\al=\al_1,\al_2, \al_y,\al_\lambda$,
having defined $\al_y=y_t^2/(4\pi)$ and $\al_\lambda=\lambda/(4\pi)$,  and
the exponentiated running factors follow by integration of the corresponding
anomalous dimension
\begin{equation}\label{eq:A}
{\cal A}_X^{(\al)} (\mu_l, \mu_h)=
 \exp\left\{-\int_{\mu_l}^{\mu_h}
 \frac{ \gamma^{(\al)}_X(\mu) }{\mu}{\rm d}\mu\right\}.
\end{equation}
Here $\ga^{(\al)}_X$ is the term proportional to the coupling $\al$ in the anomalous dimension for the Wilson
coefficient $C_X$.
By inspection of Eq.(\ref{eq:ga_tot}), we see that such terms can be written as  functions of
$t=\log\mu$
\begin{equation}
\ga^{(\al)}_X=\al \left[ a_\al t+{\cal F}^{(\al)}_X(p)\right]
\end{equation}
where for each coupling $\al$ there is a different coefficient $a_\al$,  while  ${\cal F}^{(\al)}_X(p)$  is a  function
of the external momenta $p\equiv (p_1,p_2,p_3,p_4)$  that in general also depends on the chosen coefficient $C_X$. 
The explicit expressions of the functions are
\begin{align}  \label{eq:F}
{\cal F}^{(1)}_X(p)=&\frac{1}{\pi}\sum_i Y_i^2\log(\bn_i \cdot p_i) 
-\frac{3}{4\pi}\left[ \sum_i Y_i^2 + \frac{1}{2}\right]    \,, \nn \\
{\cal F}^{(2)}_X(p)=&\frac{3}{4\pi}\sum_i \eta_i \log(\bn_i \cdot p_i)   
- \frac{9}{16\pi}\left[ \sum_i \eta_i  + d^\op \right] \,, \nn \\
{\cal F}^{(y)}_X(p)=&\frac{3}{4\pi}   \,, \nn \\
{\cal F}^{(\lambda)}_X(p)=&\frac{c^\op}{\pi}   \,, 
\end{align}
while the coefficients are
\begin{align}  \label{eq:a}
a_1=& -\frac{1}{\pi}\sum_i Y_i^2\,, \nn \\
a_2=&-\frac{3}{4\pi}\sum_i \eta_i \,,\nn \\
a_\lambda=&0 \,,\nn \\
a_y=&0 \,.
\end{align}
In all these definitions the sums are over the external quarks.
The couplings also have an implicit dependence on $t$, dictated by  the one-loop beta-functions  
\begin{equation}\label{eq:beta}
\frac{ {\rm d} \al_i}{{\rm d} t}= b_i \al_i^2  \,.
\end{equation}
These can be easily integrated, yielding
\begin{equation}\label{eq:al}
\left(\frac{\al_i(t_l)}{\al_i (t_h)}\right)=1- b_i \al_i(t_l) (t_h-t_l) 
\end{equation}
where the one-loop coefficients are well known\cite{Arason}
\begin{align}  \label{eq:b}
b_1=& \frac{41}{12\pi} \,, \nn \\
b_2=&- \frac{19}{12\pi} \,, \nn \\
b_\lambda=& \frac{3}{\pi} \,, \nn \\
b_y=& \frac{9}{4\pi} \,.
\end{align}

By a simple integration,  the running factors in  Eq.(\ref{eq:A}) take the explicit general form
\begin{align}\label{eq:Aint}
{\cal A}_X^{(\al)}& (\mu_l, \mu_h)=
\left(\frac{\mu_h}{\mu_l}\right)^{\displaystyle{ \frac{a_\al}{ b_\al} }}\nn \\ 
&\times 
\left(\frac{\al(t_l)}{\al(t_h)}\right)^{\displaystyle{   \left[\frac{1}{ b_\al}
\left( \frac{a_\al}{ b_\al \al(t_l)}+ a_\al t_l+ {\cal F}_X^{(\al)} (p)\right)\right]}}
\,
\end{align}
and their product in Eq.(\ref{eq:WC}) gives an analytical expression for the electroweak radiative correction
at LL order. The cross section follows by insertion of the low scale Wilson coefficients from Eq.(\ref{eq:WC})
in the phase space integral of Eq.(\ref{eq:si_ew}) and integrating by PDFs.
The closed analytical form of Eq.(\ref{eq:Aint}) greatly speeds up the numerical integration of the cross section
at LL order compared with NLL order. We discuss the accuracy of the two orders  
in the next section by a direct comparison.

\section{LL  vs. NLL order: numerical results }
\label{sec:numerical}

In this section the accuracy of the approximation is tested by a comparison between LL and NLL orders. All the details
of the numerical integration are kept exactly the same as reported in Ref.~\cite{Siringo:2012} where the cross
section was evaluated at NLL order. We summarize them briefly.
The low energy matching scale is chosen to be $\mu_l=M_Z$.
The couplings and parameters of the standard model have been set at the electroweak scale $\mu_l=M_Z$ according
to the data of Ref.\cite{PDG}.  The Higgs mass is assumed to be $M_H=125$ GeV.
The high scale $\mu_h$ is set at the larger value between  $M_Z$ and the geometric average $\mu^2=\sqrt{t_3 t_4}$.
At this scale the sum of logarithmic terms $\ln ({-t_3}/\mu^2)+\ln ({-t_4}/\mu^2)$ reaches its minimum in the one-loop
matching. As discussed in Ref.~\cite{Siringo:2012} this choice has the merit of stopping the running whenever one of the Mandelstam variables is too small, while keeping the neglected one-loop matching terms as small as possible.
On the other hand the sensitivity  to the choice of the high scale was shown to be small and comparable
to the sensitivity of the standard tree-level cross section.~\footnote{At variance with variational approaches the principle of minimal sensitivity does not hold. The best choice of $\mu_h$ should make the omitted terms small, as for the method of minimal variance\cite{Siringo:2005}.} 
We use CTEQ6 PDFs\cite{PDF} and neglect the very small contribution of t and b quarks. Moreover 
we neglect the s-channel contribution and interference terms that are known to be small with VBF cuts.
The masses of the vector gauge bosons are restored in the denominators of the coefficients in \eq{c_def}, as  their effects  become important when the total cross section is evaluated.
Hereafter, in order to compare with the NLL calculation of Ref.~\cite{Siringo:2012}, we adopt the same
cuts on angles and transverse moment: $\theta_3> 10^o$, $\theta_4< 170^o$, $p_T>20$ GeV.
No QCD corrections have been included in both calculations, and only the
virtual electroweak corrections are considered by the method.

The terms contributing to the cross section at tree-level 
are reported in Fig.2. The $ud\to du$
process dominates, followed by the other Left-Left terms. The Left-Right contributions are two order of
magnitude smaller, while the Right-Right terms are very small and have been neglected.
In Fig.2  a sum over the  generations of quarks is included through the PDFs.    

\begin{figure}[t] \label{fig:terms}
\centering
\includegraphics[width=0.37\textwidth,angle=-90]{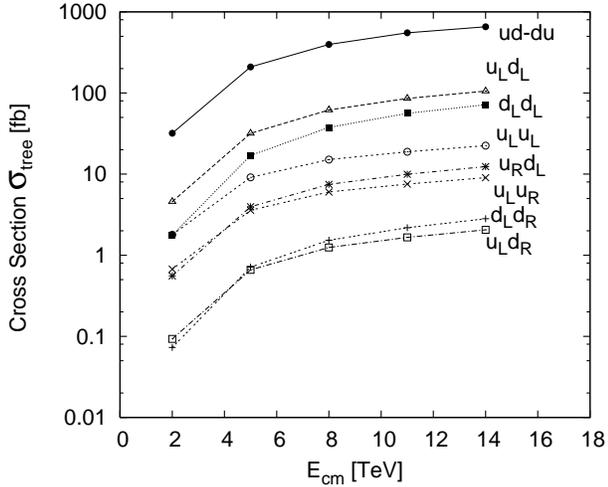}
\caption{Terms contributing to the cross section at tree-level, as a function of $\Ecm$, with
the cuts: $\theta_3> 10^o$, $\theta_4< 170^o$, $p_T>20$ GeV.}
\end{figure}

\begin{figure}[t] \label{fig:uddu}
\centering
\includegraphics[width=0.32\textwidth,angle=-90]{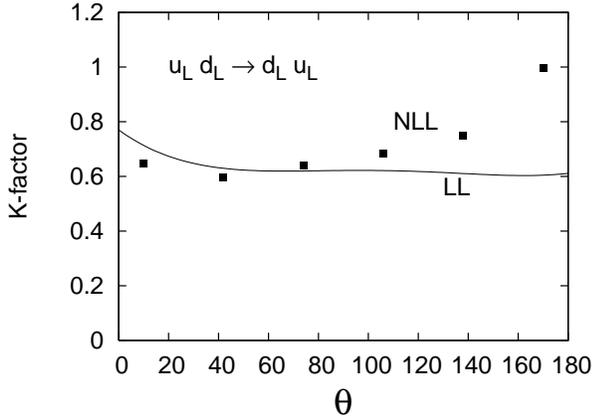}
\caption{The factor $K=\sigma_{EW}/\sigma_{tree}$ for the process $u_L d_L\to d_L u_L$
as a function of $\theta=\theta_3=180^o-\theta_4$,
at $\varphi=90^o$, $x_1=x_2=0.5$, $\Ecm=14$ TeV and $E_3=x_1 x_2\Ecm/2$. The result at LL order
(solid line) is compared with the  NLL order data points (squares).}
\end{figure}

\begin{figure}[t] \label{fig:udll}
\centering
\includegraphics[width=0.32\textwidth,angle=-90]{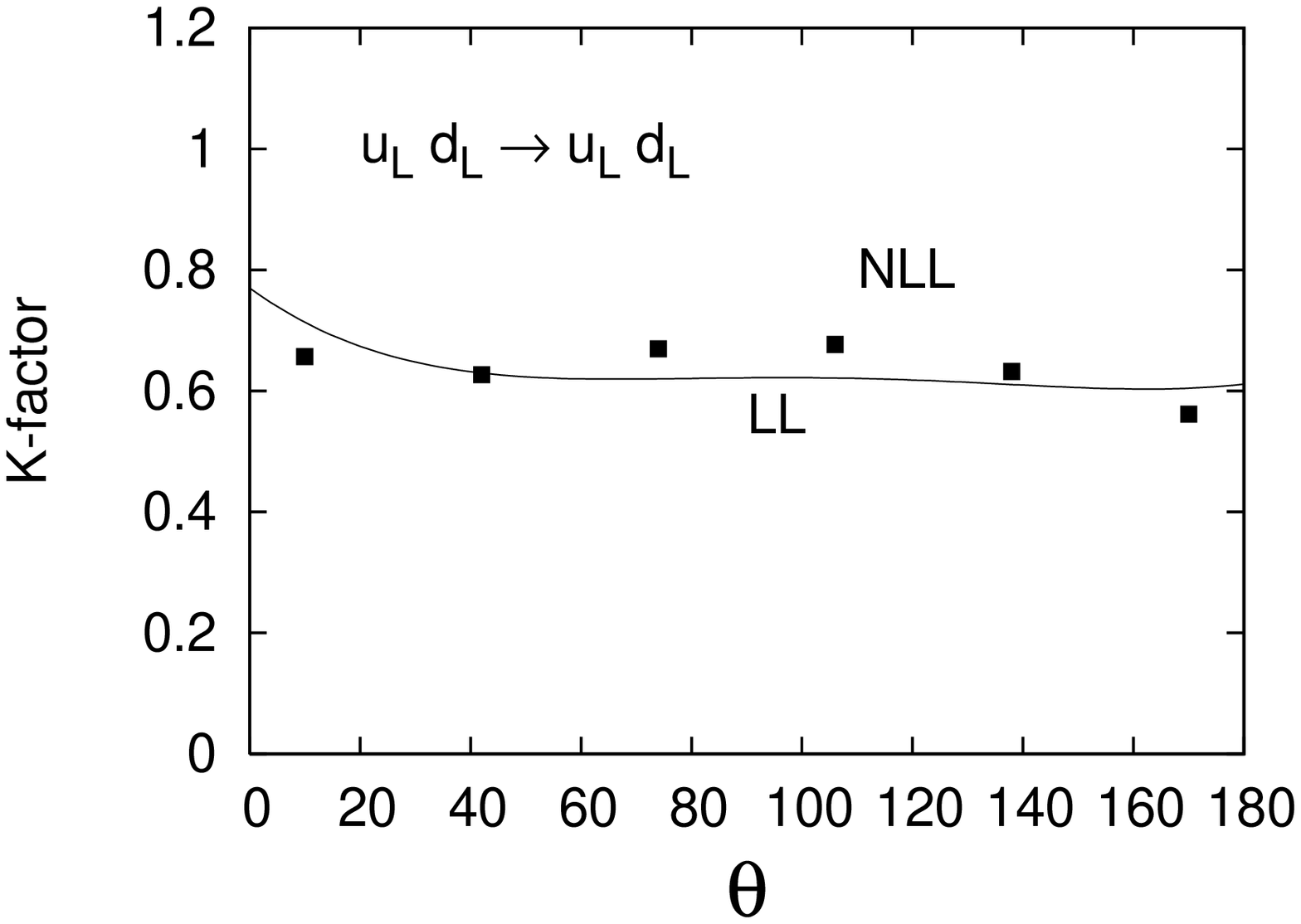}
\caption{The same as Fig.3  for the process $u_L d_L\to u_L d_L$.}
\end{figure}

\begin{figure}[!ht] \label{fig:ddll}
\centering
\includegraphics[width=0.32\textwidth,angle=-90]{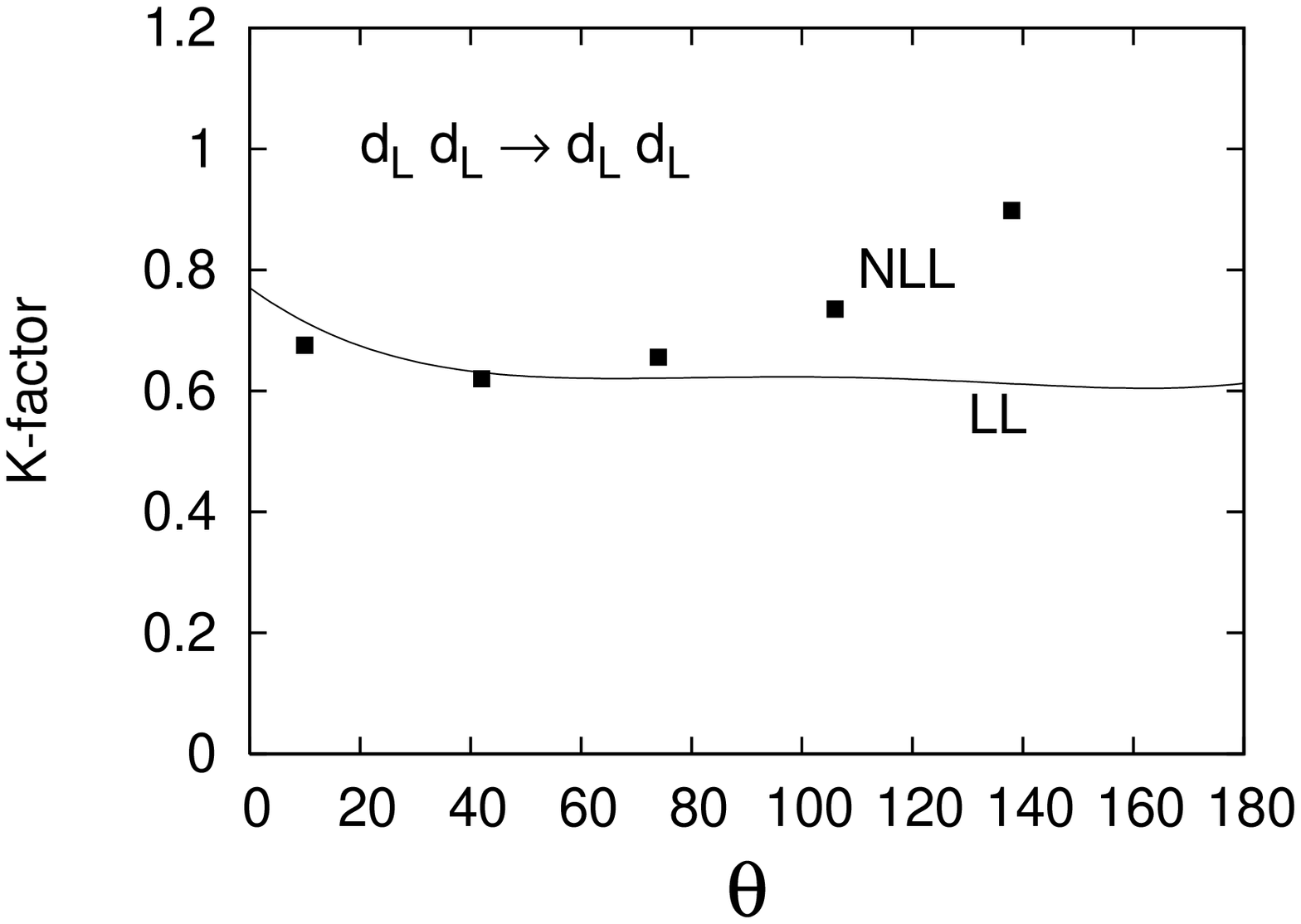}
\caption{The same as Fig.3  for the process $d_L d_L\to d_L d_L$.}
\end{figure}

\begin{figure}[htb] \label{fig:uull}
\centering
\includegraphics[width=0.32\textwidth,angle=-90]{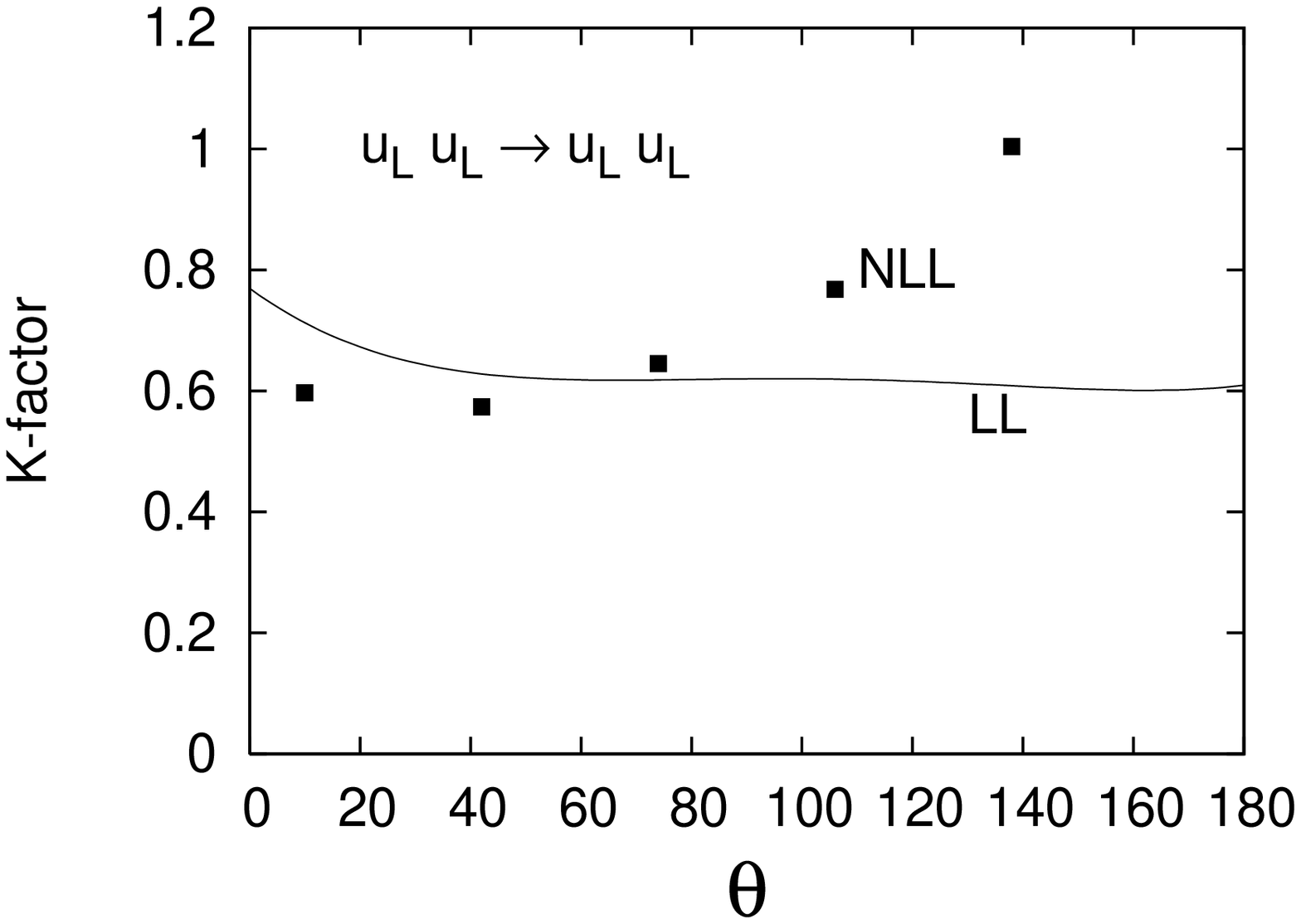}
\caption{The same as Fig.3  for the process $u_L u_L\to u_L u_L$.}
\end{figure}

\begin{figure}[t] \label{fig:udrl}
\centering
\includegraphics[width=0.32\textwidth,angle=-90]{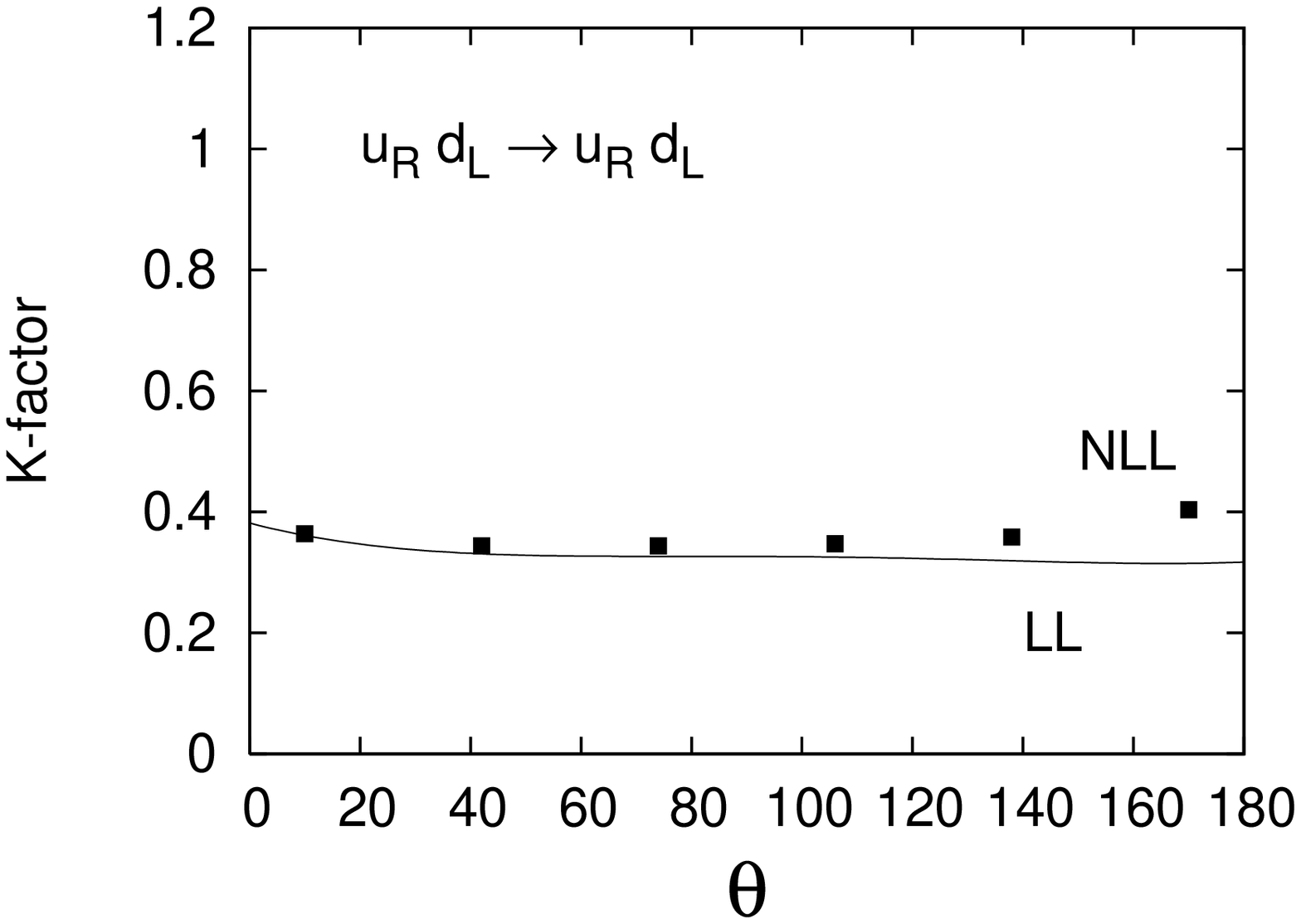}
\caption{The same as Fig.3  for the process $u_R d_L\to u_R d_L$.}
\end{figure}

As a first comparison we study the single terms in the phase space, and for each of them we calculate the
K-factor which is defined as the ratio between the cross section with electroweak corrections on, and the
tree-level cross section without any radiative correction
\begin{equation} \label{eq:kfac}
K=
\frac{\sigma_{EW}(\theta_3,\theta_4,\varphi)}{\sigma_{tree}(\theta_3,\theta_4,\varphi)}
\,,
\end{equation}
where the differential cross sections are evaluated for fixed values of $x_1$,$x_2$, $E_3$ and $\Ecm$,
and for a given set of angles. Here the simple LL calculation is compared with the NLL result of Ref.~\cite{Siringo:2012}.
In Figs. 3-10 the K-factor is reported
as a function of $\theta=\theta_3$ with $\theta_4=\pi-\theta_3$ and $\varphi=\pi/2$ at a typical
set of parameters: $\Ecm=14$ TeV, $x_1=x_2=0.5$ and $E_3=x_1x_2\Ecm/2$.  The 
agreement of  LL and NLL results is very good at small angles. As expected, at LL order the dependence on angles is reduced in comparison with  the NLL result, because the neglected terms have a larger dependence on the scattering angles. However for $\theta<90^o$,  in the  physically relevant phase-space region, the difference is small and the
LL result (solid line) interpolate the NLL data (squares) quite well. The dependence on the azimuthal angle $\phi$
is very small and negligible for $\theta<90^o$,  in perfect agreement with the  NLL order.

\begin{figure}[!ht] \label{fig:uulr}
\centering
\includegraphics[width=0.32\textwidth,angle=-90]{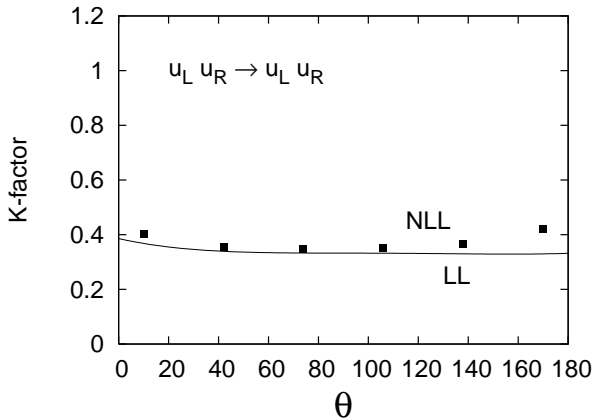}
\caption{The same as Fig.3  for the process $u_L u_R\to u_L u_R$.}
\end{figure}

\begin{figure}[!ht] \label{fig:ddlr}
\centering
\includegraphics[width=0.32\textwidth,angle=-90]{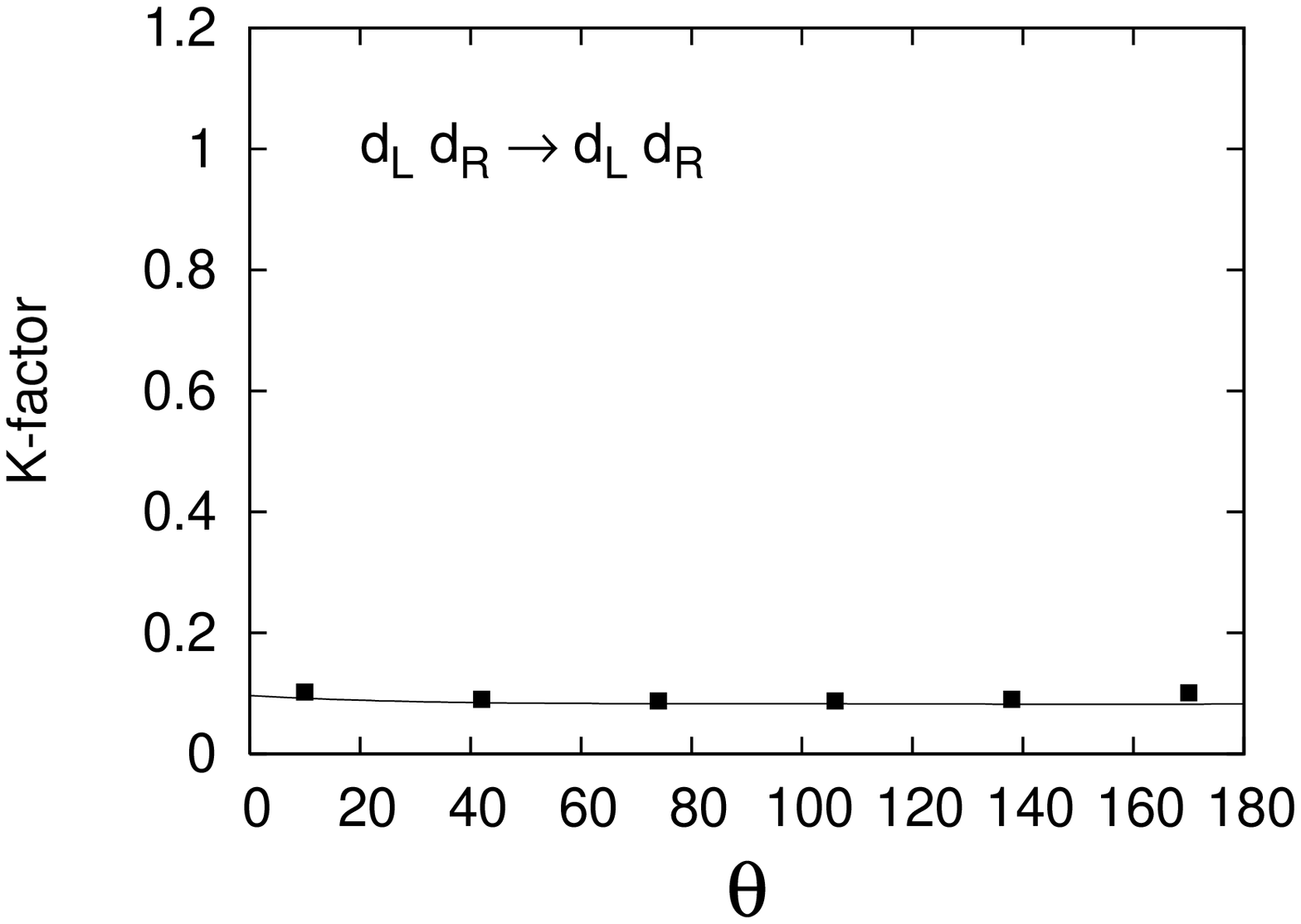}
\caption{The same as Fig.3  for the process $d_L d_R\to d_L d_R$.}
\end{figure}

\begin{figure}[!ht] \label{fig:udlr}
\centering
\includegraphics[width=0.32\textwidth,angle=-90]{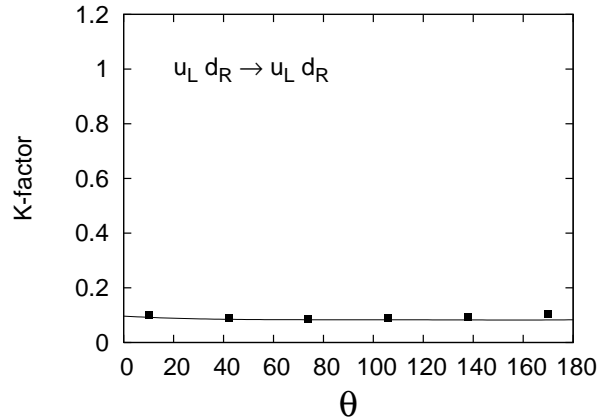}
\caption{The same as Fig.3  for the process $u_L d_R\to u_L d_R$.}
\end{figure}

We find a better agreement for the Left-Right processes  (Fig. 7-10) where the radiative corrections are
larger, than for Left-Left processes (Fig. 3-6) where the corrections are smaller. Thus the very large suppression
of the Left-Right processes is mainly due to the role played by the collinear anomalous dimension.
It is remarkable that such  large suppression  has no
relevant effects on the total cross section which is dominated by the Left-Left processes.

The single terms are integrated by PDFs over the phase space and summed up, yielding
the cross section $\sigma_{EW}$. 
By comparison with the tree level cross section,  the relative
electroweak correction  is defined as the ratio
\begin{equation} \label{eq:relsig}
\frac{\Delta\sigma}{\sigma}=
\frac{\sigma_{EW}-\sigma_{tree}}{\sigma_{tree}}
\,,
\end{equation}
and is displayed in Fig.11  
as a function of the center-of-mass energy. For comparison the NLL result of  Ref.~\cite{Siringo:2012}
is also reported in Fig.11 together with the output of 
the code HAWK\cite{Ciccolini:2007jr,Ciccolini:2007ec} that is based on a 
fixed order perturbative calculation.

\begin{figure}[t] \label{fig:ewcorr}
\centering
\includegraphics[width=0.32\textwidth,angle=-90]{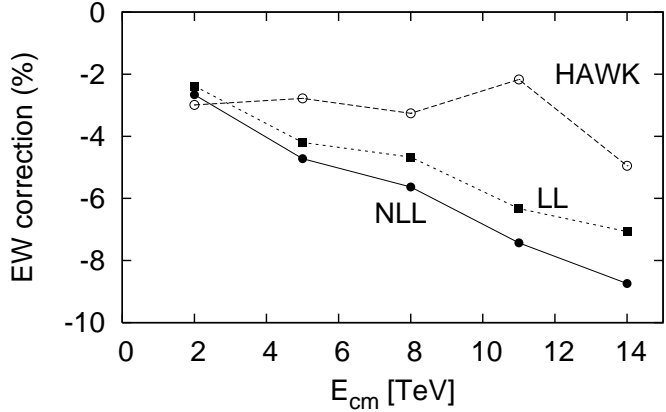}
\caption{Relative electroweak correction $[\sigma_{EW}-\sigma_{TREE}]/\sigma_{tree}$
as a function of the center-of-mass energy $\Ecm$ evaluated by \eq{si_ew} at LL
order (filled squares), compared with the NLL order of Ref.~\cite{Siringo:2012} (filled circles) and
with the output of HAWK (open circles).
Cuts are: $p_T>20$ GeV, $\theta_3> 10^o$, $\theta_4< 170^o$.}
\end{figure}

While differences are negligible below 2 TeV,  at large energies   LL and NLL corrections grow
faster than predicted by  fixed order one-loop calculations, reaching  9\% at the
full LHC energy $\Ecm=14$ TeV, to be compared with 5\% predicted by HAWK.
At LL order the correction is a bit smaller with respect to NLL order, but the difference
is less than 1\% up to $\Ecm=10$ TeV. Thus the simple LL calculation provides an analytical
electroweak correction that seems to be more reliable than fixed order calculations at high energies.
While fixed-order perturbative calculations might miss part of the correction at the LHC energy scale,
the simple LL resummation contains the main terms, and gives integrated corrections that are  more accurate than  PDFs.
Insertion of the simple analytical result of Eqs.(\ref{eq:WC}),(\ref{eq:Aint}) 
in software packages would be straightforward, and would increase
the accuracy of simulations.

\end{document}